\documentclass[aps,prb,floatfix,preprint]{revtex4}
\usepackage{graphicx}
\usepackage{amsmath}
\usepackage[dvips]{epsfig}
\usepackage{inputenc,exscale,amsmath,amssymb,epsfig,rotating,psfrag,bbm,color}
\usepackage{layout}
\usepackage{calc}
\usepackage{tabularx}
\usepackage{pstricks,pst-node,pst-tree}
\usepackage{array,multirow,bigdelim,bm}

\def\he#1{$\rm^{#1}He $}

\def\rvec{{\bf r}}

\begin{document}

\title{Solvation of Mg in Helium-4: Are there Meta-stable Mg Dimers?}

\author{Eckhard Krotscheck$^{1,2}$\renewcommand{\thefootnote}
{\alph{footnote}}\footnote{Electronic address: eckhardk@buffalo.edu},
Robert E. Zillich$^2$\renewcommand{\thefootnote}
{\alph{footnote}}\footnote{Electronic address: robert.zillich@jku.at}}
\affiliation{$^1$Department of Physics, University at Buffalo SUNY,
Buffalo NY 14260, USA\\
$^2$Institute for Theoretical Physics, Johannes Kepler
  Universit\"at Linz, Altenbergerstra\ss e 69, A-4040 Linz, Austria}

\date{\today}

\begin{abstract}
  Experiments on the formation of magnesium complexes in $^4$He
  nanodroplets were interpreted as the observation of the formation of
  weakly bound magnesium complexes. We present results for single Mg
  and Mg dimer solvation using the hypernetted chain / Euler-Lagrange
  method as well as path integral Monte Carlo simulations.  We find
  that the phonon-mediated, indirect Mg-Mg interaction adds an
  oscillatory component to the direct Mg-Mg interaction.  We undertake
  a step-by-step examination of the ingredients of the calculation of
  the phonon-induced interaction, comparing the results of
  semi-analytic HNC-EL calculations for bulk and single impurity
  results with experiments as well as Monte Carlo data.  We do not
  find evidence for a sufficiently strong secondary minimum in the
  effective Mg-Mg interaction to support a metastable state.
\end{abstract}

\maketitle

\section{Introduction}

Experiments by Przystawik {\it et al.\/} \cite{mgfoam_expt} on the
formation of magnesium complexes in $^4$He nanodroplets were
interpreted as the observation of the formation of weakly bound
magnesium complexes, dubbed ``bubble foam'', which collapsed to form
dense Mg clusters upon laser excitation.  In the same work, as well as
in Ref.~\onlinecite{hernandoPRB08b}, calculations based on a density
functional theory (DFT) for helium found that the effective
interaction potential between two Mg atoms has a local minimum at a
separation of about 10\AA and 2-3~K deep.  The presence of such a
minimum would support the interpretation that indeed a weakly bound
``bubble foam'' can form in helium where Mg atoms are separated from
each other by a layer of $^4$He atoms. The mean lifetime of a
metastable dimer state was, however, estimated to be many orders of
magnitude too short to explain the experiment. Such a predicted dimer
could only in a highly excited rotational state be stable on the time
scale of the experiment.

We examine in this work the effective interaction potential between
two Mg atoms solvated in $^4$He using both a semi-analytic approach
based on optimized pair and triplet correlations quantum Monte Carlo
simulations.  We expect that, if weakly bound Mg dimers can form, this
would be a local effect and not be affected by the presence of the
droplet surface.  That means that such an effect would be observable
in small and large $^4$He droplets as well as in bulk $^4$He. To
determine if such a bound state can exist, we perform calculations of
Mg impurities in bulk $^4$He using the hypernetted chain /
Euler-Lagrange (HNC-EL) method.  HNC-EL yields the effective potential
$V_{\rm eff}(r)$ felt by two Mg atoms inside helium which consist of
the bare interaction and an oscillatory induced potential mediated by
phonons in the surrounding He.  We show that these oscillations are
not strong enough to lead to pronounced secondary minima of $V_{\rm
  eff(r)}$.  We have carefully examined hydrodynamic consistency of all
ingredients, and the importance of so-called elementary diagrams and
triplet correlations and find these negligible at distances where a
metastable state would exist.

As far as computationally efficient, we compare our HNC-EL results for
single Mg in bulk helium  with corresponding path integral Monte Carlo
(PIMC) simulations where we found excellent agreement.  Furthermore we
have performed simulations of Mg  dimers and trimers in small droplets
of up to 100 $^4$He atoms.  Again, we did not observe the formation of
weakly bound Mg  complexes but instead observed  a swift equilibration
to the ground state.

\section{Methodology}

In this work we employ two complementary methods, a semi-analytic
approach using diagrammatic quantum many-body theory, and a
computational approach using quantum Monte Carlo simulation.  Both
methods employ the full many-body Hamiltonian of $N$ $^4$He atoms
(indexed by Latin subscripts) and one or more Mg impurities (indexed
by Greek subscripts)
\begin{equation}
\label{eq:H}
  H = -\frac{\hbar^2}{2M}\sum_\alpha\nabla_\alpha^2
  -\frac{\hbar^2}{2m}\sum_{i=1}^N\nabla_i^2\
    +\ \sum_{\alpha,i}V_{AHe}(|\rvec_\alpha-\rvec_i|)
    + \sum_{i<j}V_{HeHe}(|\rvec_i-\rvec_j|)
\end{equation}
where $m$ and $M$ are the masses of a $^4$He and Mg atom,
respectively. In this Hamiltonian we assume pair-wise interactions
which is known to be a reliable approximation for atoms in their
electronic ground state. We assume that the \he4 atoms interact via
the Aziz-II potential \cite{AzizII}. This potential reproduces the
equation of state of \he4 with high accuracy \cite{JordiQFSBook}. For
the interaction $V_{AHe}(\rvec_\alpha-\rvec_i)$ between the host
liquid and the impurity atom we have used the potential models by
Hinde~\cite{Hinde03} {\it et al.\/} and, for comparison, the one by
Partridge {\it et al.\/}~\cite{partridgeJCP01}.  For the Mg-Mg
interaction we have used the potential by Tiesinga {\it et
  al.\/}~\cite{tiesingaPRA02}.

\subsection{Hypernetted-chain Euler-Lagrange}
\label{ssec:HNCEL}

\subsubsection{\he4 background calculation}

The Hypernetted chain/Euler-Lagrange method is a well established fast
and accurate method for calculating properties of strongly interacting
quantum fluids.  The method has been described in numerous review
articles and pedagogical material, see, for example,
Ref. \onlinecite{MikkoQFSBook}. The wave function is expanded in a
Jastrow-Feenberg form in terms of multiparticle correlation functions
$u_n(\rvec_1,\ldots,\rvec_n)$, truncation at $n=3$ is normally
sufficient.
\begin{equation}
        \Psi_0({\bf r}_1,\ldots,{\bf r}_N) =
        \exp\frac{1}{2}\Bigl[
         \sum_{i<j} u_2({\bf r}_i,{\bf r}_j)
        + \sum_{i<j<k} u_3({\bf r}_i,{\bf r}_j,{\bf r}_k)
        + \ldots\Bigr].
\label{eq:Jastrow}
\end{equation}
The correlation functions $u_n({\bf r}_1,\ldots,{\bf r_n})$ are
determined by minimization of the energy-expectation value $E_0$
\begin{equation}
        \frac{\delta E_0}{\delta u_n({\bf r},\ldots,{\bf r_n})} = 0.
\label{eq:euler}
\end{equation}
The energy expectation value and other physically relevant quantities
are calculated by diagrammatic expansions.  The hierarchy of
``hypernetted chain'' integral equations provides a scheme that is, at
every level of implementation, consistent with the optimization
problem (\ref{eq:euler}) in the sense that the resulting pair
distribution and structure functions reproduce {\it qualitatively\/}
the properties of the solution of the exact variational problem.

For the problem at hand, we only need the static structure function
$S(k)$ for bulk helium.  For long wavelengths, the static structure
function goes as
\begin{equation}
  S(q) = \frac{\hbar q}{2mc}\quad\mbox{as}\quad q\rightarrow 0
\label{eq:Slong}
\end{equation}
where $c$ is the speed of sound obtained from a theory of excitations
or from the equation of state
\begin{equation}
mc^2 = \frac{d}{d\rho}\rho^2\frac{d (E/N)}{d\rho}\,.
\label{eq:mceos}
\end{equation}

The $c$ obtained from the equation of state will normally be {\it
  different\/} from the speed of sound obtained from the slope of
$S(q)$, in fact these two quantities agree only in an exact theory.
However, the inconsistency is weak and can be repaired by a slight
phenomenological modification of the Euler equation for the triplet
correlation function which has no visible consequences on the equation
of state \cite{lowdens}.

\subsubsection{Singe impurity calculation}

In the next step, the impurity is included. The wave function of the
compound system is
\begin{eqnarray}
        \Psi^I (\rvec_0,\rvec_1,...,\rvec_N)
        = \exp \frac{1}{2} \Bigl[ \sum_{j=1}^N u^{I}(\rvec_0,\rvec_j)
        +\frac{1}{2!}\mathop{\sum_{j,k=1}}_{j\ne k}^N
        u^{I}(\rvec_0,\rvec_j,\rvec_k)
        \Bigr]\Psi(\rvec_1,...,\rvec_N).
\nonumber\\
\label{eq:waveI}
\end{eqnarray}
The {\it chemical potential\/} of the impurity is the energy gained or
lost by adding one impurity particle into the liquid, in other words it
equals to the energy difference
\begin{eqnarray}
        \mu^I &=& E_{N+1} - E_N
\nonumber\\
        &=& \frac{\langle\Psi^I\vert H^I \vert \Psi^I\rangle
                }{ \langle\Psi^I\vert\Psi^I\rangle}
        -\frac{\langle\Psi\vert H \vert \Psi\rangle}{
        \langle\Psi\vert\Psi\rangle}.
\label{eq:chemi}
\end{eqnarray}
and the Euler equation are derived by minimizing the impurity chemical
potential. Results are again the energetics as well as structure and
distribution functions.

There is again a hydrodynamic consistency condition between the long
wavelength limit of the static structure function and the density
dependence of the chemical potential: The long wavelength limit of the
impurity--background (``IB'') structure function is \cite{FeenbergBook}
\begin{equation}
S_{IB}(q) = -\alpha \quad\mbox{as}\quad q\rightarrow 0
\label{eq:alphaSofq}
\end{equation}
where $\alpha$ is the volume excess factor
\begin{equation}
\alpha = \frac{\rho}{mc^2}\frac{d \mu_{I}}{d\rho}\,.
\label{eq:alphaEOS}
\end{equation}
The volume excess factor $\alpha$ can be calculated from $S_{IB}(0+)$
and from the density dependence of the impurity binding energy. Again,
the long wavelength limit (\ref{eq:alphaSofq}) and the hydrodynamic
derivative (\ref{eq:alphaEOS}) agree only in an exact theory.  By the
same slight phenomenological modification of the three-body
correlation function that enforces the consistency between the speed
of sound obtained by Eqs. (\ref{eq:Slong}) and (\ref{eq:mceos}) and
does not affect the impurity binding energy visibly, the two
quantities can be made consistent.

\subsubsection{Impurity-impurity interaction}

In the case of two impurities, located at $\rvec_0$ and $\rvec_0'$,
the wave function is
\begin{eqnarray}
        \Psi^{II} (\rvec_0,\rvec_0';\rvec_1,...,\rvec_N)
        &=& \exp \frac{1}{2} \Bigl[ u^{II}(\rvec_0,\rvec_0')
        + \sum_{j=1}^N \left[u^{IB}(\rvec_0,\rvec_j)+ u^{IB}(\rvec_0',\rvec_j)
          + u^{IIB}(\rvec_0,\rvec_0',\rvec_j)
          \right]\nonumber\\
          &&\quad
        +\frac{1}{2!}\mathop{\sum_{j,k=1}}_{j\ne k}^N\left[
        u^{IBB}(\rvec_0,\rvec_j,\rvec_k)+u^{IBB}(\rvec_0',\rvec_j,\rvec_k)\right]
        \Bigr]\Psi_0(\rvec_1,...,\rvec_N)\,.
\label{eq:waveII}
\end{eqnarray}
The situation for the two impurity case differs from the above ones
because the wave function (\ref{eq:waveII}) can only lead to an
effective impurity-impurity potential determining the
impurity--impurity correlation function
$u^{II}(\rvec_0,\rvec_0')$. This is exactly the quantity we want
because it gives us the information on the configuration.  Its form
has, in HNC approximation, been first derived by Owen \cite{Owen};
adding ``elementary diagram'' and ``triplet correlation'' corrections,
the effective interaction is
\begin{equation}
V_{eff}(r) = V^{(II)}(r) + V_e(r) + w_{\rm ind}(r)
\label{eq:Veff}
\end{equation}
where $V^{(II)}(r)$ is the bare interaction between the two
impurities, $V_e(r)$ the correction from ``elementary'' diagrams and
triplets, and $w_{\rm ind}(r)$ is the induced interaction originating
from phonon exchange and higher--order processes. The induced
potential can depend only on background and single-impurity
quantities. The ``elementary diagram'' correction is calculated within
the usual diagrammatic expansion methods \cite{ekthree}. The induced
potential from triplet calculations is new; it is best calculated
taking the two-impurity limit of the mixture
theory\cite{mixmonster}. Since triplet correlations turn out to cause
a negligible correction we refrain from giving the somewhat tedious
derivation.

Jastrow-Feenberg theory provides a prescription for calculating the
induced Mg-Mg potential mediated by the density fluctuations in the He
liquid, {\it i.e.\/} phonon exchange \cite{Owen}
\begin{equation}
  \tilde w_{\rm ind}(k) = \tilde V_{\rm ind}(k,\bar\omega(k))
  = -\frac{\hbar^2 k^2}{4m}
  \left[\frac{S_{IB}(k)}{S(k)}\right]^2
  \left[2\frac{m}{m_I}S(k)+1\right]\,.
\label{eq:wind1}
\end{equation}

One can interpret this term from linear response theory: The full
interaction between two Mg atoms is strictly speaking energy
dependent, it is the sum of the induced and the bare interaction,
\begin{equation}
V_{\rm eff}(r,\omega) = V_{\rm Mg-Mg}(r) + V_{\rm ind}(r,\omega)
\label{eq:veff}
\end{equation}
where
\begin{equation}
\tilde V_{\rm ind}(k,\omega) = \tilde V_{IB}(k)\chi(k,\omega)\tilde V_{IB}(k)
+\tilde V_e(k)
\end{equation}
and $\chi(k,\omega)$ is the density-density response function of bulk
helium, $V_{ph}^{IB}$ is the particle-hole potential, and $\tilde
V_e(k)$ consists of contributions from triplets and elementary
diagrams.  The HNC-EL result is obtained by taking the density-density
response function at an {\em average\/} imaginary
frequency\cite{mixmass} that is chosen according to the localization
rules of parquet-diagram
theory\cite{parquet1,parquet2,parquet3,parquet4}.

If one looks at the phonon exchange of weakly bound pairs of Mg atoms,
one might therefore argue that it is better to take $\omega=0$ which
leads to a slightly different effective interaction
\begin{equation}
  \tilde w_{\rm ind}'(k) = \tilde V_{\rm ind}(k,\omega=0)
  = -\frac{\hbar^2 k^2}{4m}
  \left[\frac{S_{IB}(k)}{S(k)}\right]^2
  \left[\phantom{2}\frac{m}{m_I}S(k)+1\right]\,.
\label{eq:wind2}
\end{equation}

We will see, however, that there is little difference between the
predictions of these two procedures for calculating the induced
interaction.

\subsection{Path integral Monte Carlo}
\label{ssec:PIMC}

Path integral Monte Carlo (PIMC) exploits the isomorphism between quantum theory and a classical
system of closed polymer chains~\cite{feynmanhibbs,chandlerJCP81}. For
bosons, like $^4$He atoms, PIMC is an exact finite-temperature method.
In PIMC the configuration space representation of the many-body
density matrix $\rho(\rvec,\rvec';\beta) = \langle \rvec|e^{-\beta
  \hat H}|\rvec'\rangle$, where $\beta=1/k_\text{B}T$, is sampled by
the Metropolis algorithm~\cite{metro53}.  For the evaluation of
$\rho(\rvec,\rvec';\beta)$, the ``imaginary time'' interval $\beta$ is
split into smaller time steps $\tau=\beta/M$ which necessitates the
introduction of new coordinates at intermediate time slices,
\begin{equation}
\label{eq_density0M}
  \rho(\rvec_0,\rvec_M;\beta) = \int d\rvec_1\cdots\rvec_{M-1}
  \rho(\rvec_0,\rvec_1;\tau)\cdots\rho(\rvec_{M-1},\rvec_M;\tau).
\end{equation}
In the above equation, $(\rvec_0,\dots,\rvec_M)$ is a discretized path in
imaginary time.  For sufficiently small $\tau$,
$\rho(\rvec_0,\rvec_1;\tau)$ can be approximated in various ways;
here we use the pair density
approximation~\cite{ceperley95}. If only averages of diagonal operators
are required (such as energy or density distributions), we set
$\rvec_M=\rvec_0$.  Finally, Bose symmetry is implemented by
symmetrization of the density matrix
\begin{equation}
\label{eq_density_sym}
  \rho_B(\rvec,\rvec;\beta) = \frac{1}{ N!}\sum_P \rho(\rvec,P\rvec;\beta)
\end{equation}
which corresponds to reconnecting the imaginary time paths to form
larger chains.  For a detailed review of the PIMC method for bosons
see Ref.~\onlinecite{ceperley95}, for the application to dopants in
$^4$He clusters see Ref.~\onlinecite{blinovJCP04,zillichJCP05}.  Bulk
simulations are implemented by invoking periodic boundary conditions.
We put 256 $^4$He atoms together with the Mg atom in a cubic
simulation box of the side length $L$ which is determined by the
condition that the pair density $\rho_{IB}(r)/\rho_I$ approaches the
equilibrium density of bulk $^4$He, $\rho=0.02186\,$\AA$^{-3}$, for
large distance $r$.  Due to the periodic boundary conditions, the
maximal distance is typically ${L/2}\approx 11$\AA, which is not large
enough that $\rho_{IB}(r)/\rho_I$ can be considered constant.
Therefore the choice of $L$ is somewhat ambiguous. This leads to some
uncertainty in the energy calculation to be discussed below.


\section{Results}

\subsection{A single Mg impurity in bulk $^4$He}

\subsubsection{Energetics.}

\begin{figure}[hbt]
\includegraphics[width=0.6\textwidth,angle=-90]{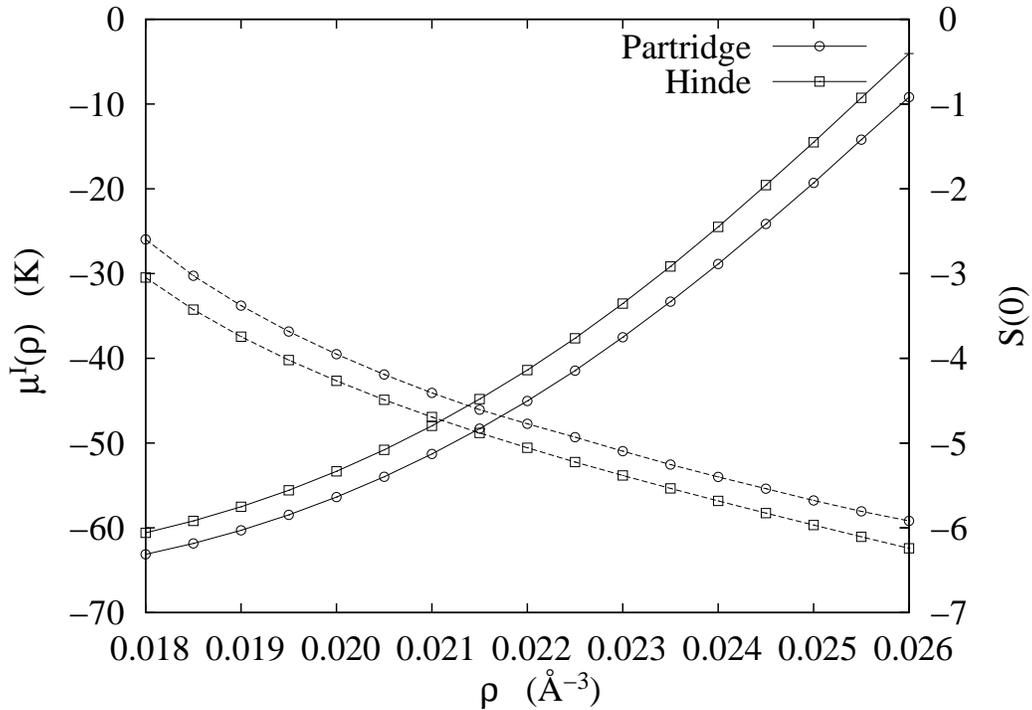}
\caption[]{\label{fig:muMG} The figure shows the chemical potential of
  a single Mg impurity in liquid \he4 as a function of density for the
  Partridge \cite{partridgeJCP01} (circles) and the Hinde\cite{Hinde03}
  potentials (boxes) (solid lines, left scale).  We also show the
  volume coefficient $-\alpha = S(0+)$ for the same two potentials
  (dashed lines, right scale).
  \label{fig:muMg}
}
\end{figure}

Fig. \ref{fig:muMg} shows HNC-EL results for the chemical potential of
a single Mg atom as a function of density, calculated for our two
potential models.
Also shown is the volume excess factor $\alpha$.

Monte Carlo calculations of the Mg chemical potential in bulk $^4$He
are computationally very expensive because the chemical potential is a
quantity of order ${\rm O}(N^0)$ and has to be extracted from the
differences of total energies which are of order ${\rm O}(N^1)$.  Due
to the large size of the Mg atom, correlated sampling as used {\em
  e.g.\/} in Ref.~\onlinecite{boronatPRB99} for an $^3$He impurity is
not possible here. HNC-EL calculations are, compared to this, many
orders of magnitude faster.  Their accuracy also suffers from the
large core of the He-Mg interaction which has the consequence that
``elementary diagrams'', which contribute only a relatively small
amount to the binding energy of \he4, are a large effect in the
problem at hand. We note, however, that the binding energy of a single
Mg impurity is only of secondary interest for our purpose; the more
relevant quantity is the static structure function $S_{IB}(k)$ and in
particular the consistency of $S_{IB}(0+)$ with the volume coefficient
$\alpha$, {\em cf.}  Eq. (\ref{eq:alphaSofq}). We have therefore
focused our attention to computing the static structure function
$S_{IB}(k)$ with PIMC which is not plagued by the difficulty of being the
difference of two big numbers.

\subsubsection{Structure.}

The most interesting quantities are the static structure function and
the pair distribution function. DMC and PIMC calculations for the
Mg-He distribution function $g_{IB}(r)$ and the static structure function
$S_{IB}(k)$ are much less demanding than energy calculations. A similar
statement applies for the HNC calculations: The largest uncertainty in
the calculation is how ``elementary diagrams'' are dealt with. While
these diagrams have a rather visible effect on the energetics, they
produce only a modest enhancement of the nearest neighbor peak in the
pair distribution function and a practically invisible correction to
the static structure function.

\begin{figure}[hbt]
\includegraphics[width=0.7\textwidth]{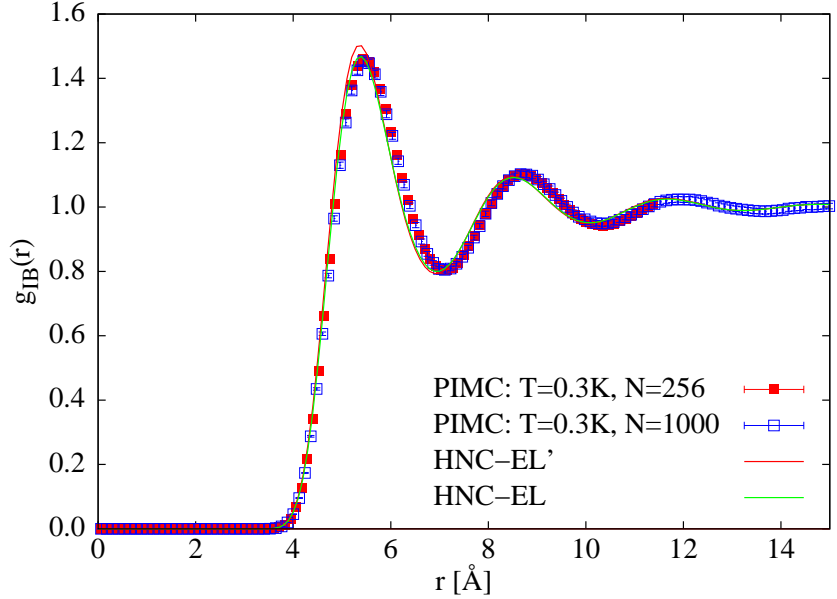}
\caption[]{\label{fig:gPIMC} Comparison of the pair distribution
  function between Mg and $^4$He in bulk helium, $g_{IB}(r)$, between
  HNC-EL and PIMC.  The HNC-EL calculation was done with and without
  enforcing hydrodynamic consistency.  The PIMC simulations were done
  at $T=0.31$K for two system sizes, $N=256$ and $N=1000$.  }
\end{figure}

\begin{figure}[hbt]
\includegraphics[width=0.7\textwidth]{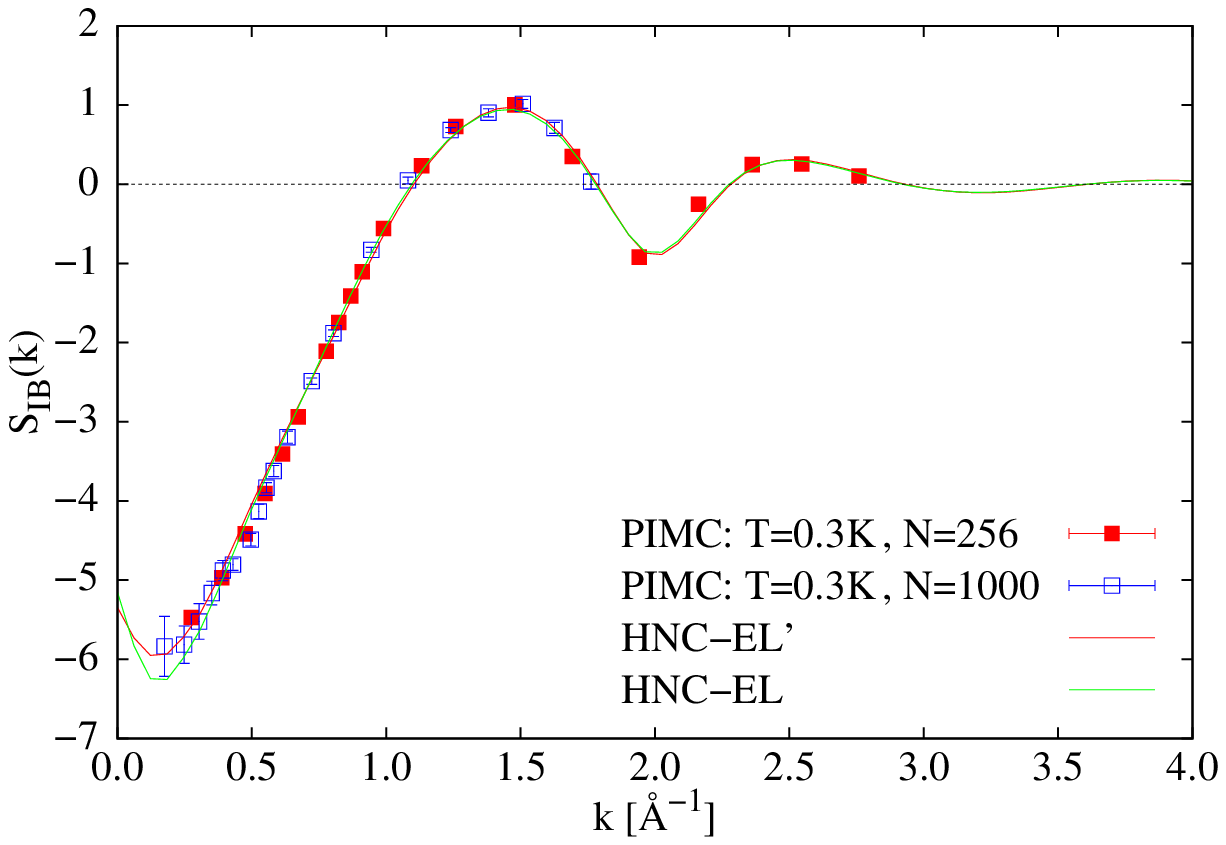}
\caption[]{\label{fig:sPIMC}
  Same as Fig.~\ref{fig:gPIMC} for the static structure function
  $S_{IB}(k)$.
}
\end{figure}

Figs. \ref{fig:gPIMC} and Figs. \ref{fig:sPIMC} show the HNC-EL and
PIMC results for the pair distribution function and the static
structure function. We find very good agreement between these results
within statistical uncertainties, confirming that both methods provide
the same high level of accuracy. As expected, the simulation results
gave somewhat larger uncertainties at long wave lengths, but the
comparison with HNC-EL calculations and the hydrodynamic consistency
condition are encouraging. Fig. \ref{fig:sPIMC} also show the
$S_{IB}(k)$ that is obtained without enforcing hydrodynamic
consistency.  Evidently the difference is small, but it appears that
the consistent $S_{IB}(k)$ agrees somewhat better with Monte Carlo
calculations.

\subsection{Single Mg impurity in $^4$He clusters}

The Mg-He interaction is quite weak, it has therefore been suggested
that Mg is not solvated inside a He cluster but might be the cluster
surface.  In a careful study using diffusion Monte Carlo, Mella {\em
  et al.}~\cite{mellaJCP05} indeed showed that the energetic balance for Mg
is very delicate, and the solvation structure indeed depends on the
details of the Mg-He interaction.  For the potential used in this
work~\cite{Hinde03}, Mella {\em et al.\/} predicted that the Mg
resides inside $^4$He$_N$ clusters at $T=0$, if $N$ is larger than
$N_{\rm crit}\approx 25$.  DFT calculations gave similar
results\cite{hernandoPRB08b}.
 
\begin{figure}[hbt]
\includegraphics[width=1\textwidth]{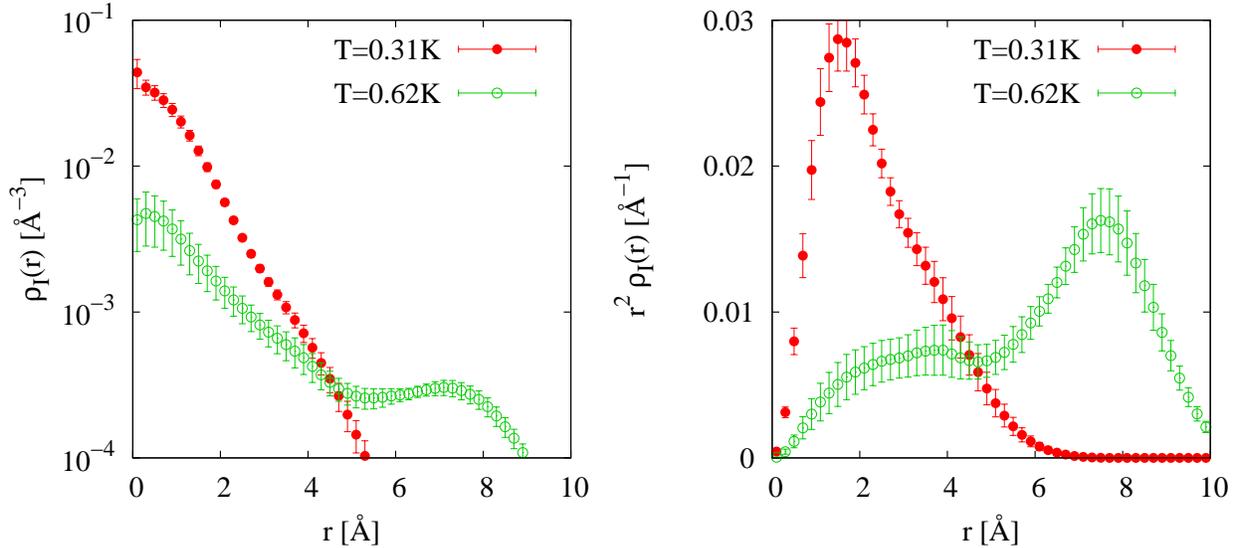}
\caption[]{\label{fig:densMg} Left: Mg density $\rho_I(r)$ with
  respect to the center of mass of the $^4$He$_N$ cluster, $N=100$, at
  $T=0.31$K and $T=0.62$K.  Right: corresponding plot $r^2\rho_I(r)$.
  The Mg impurity is clearly located inside the He cluster for
  $T=0.31$K, while for $T=0.62$K it is more likely to reside on the
  surface.}
\end{figure}

We have performed PIMC simulations of a single Mg impurity in $^4$He
clusters mainly to confirm or refute this prediction for $T>0$.  We
have chosen $T=0.31$K and $T=0.62$K, while He droplets produced in
experiment have an estimated temperature of about $T=0.3 - 0.4$K.  In
the left panel of Fig.~\ref{fig:densMg} we show the Mg density
$\rho_I(r)$ with respect to the center of mass of the $^4$He$_N$
cluster.  The probability density for Mg situated at distance $r$ from
the center mass is given by $4\pi r^2\rho_I(r)$ and is shown in the
right panel.  We chose $N=100$ to be safely in the regime where Mg is
predicted to be inside the He cluster at $T=0$.  Indeed, at the lower
end of the temperature estimated in experiments, at $T=0.31$K, the
most probable location of Mg is in the center of the He cluster.  The
situation changes drastically, as the temperature is raised to
$T=0.62$K, slightly above the experimental estimate.  The Mg density
$\rho_I(r)$ now has a shoulder at large distance for the He cluster
center.  The right panel showing $4\pi r^2\rho_I(r)$ shows that the
most probable location of Mg is now at the surface of $^4$He$_{100}$.
This strong dependence of the Mg location on the temperature around
the experimental He droplet temperature indicates that slight
disturbances of the He cluster may cause large movements of the Mg
impurity.

\newpage

\subsection{HNC-EL for the dimer interaction}

\begin{figure}[hbt]
{\includegraphics[width=0.4\textwidth,angle=-90]{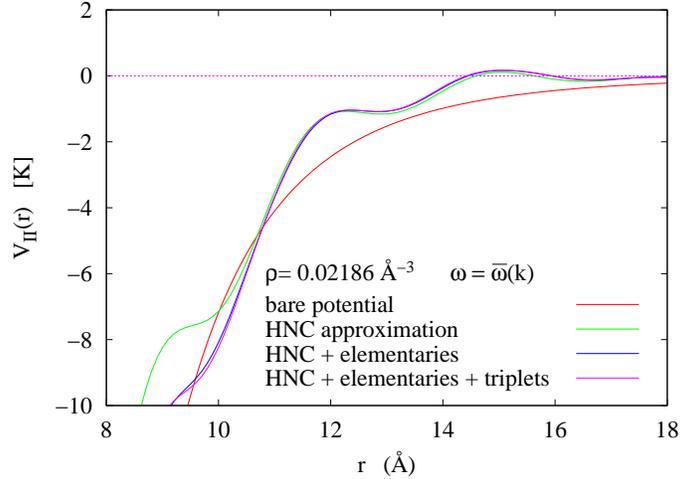}}
\caption[]{\label{fig:viiu}
  Effective Mg-Mg potential $V_{\rm eff}(r)$, consisting of the bare
  Mg-Mg interaction (full) and the induced interaction $w_{\rm ind}(r)$ given
  by Eq.~(\ref{eq:wind1}).  The three dashed lines show
  $V_{\rm eff}(r)+w_{\rm ind}(r)$ for the three increasingly accurate
  approximations: HNC-EL without elementary diagrams and triplet correlations;
  inclusion of elementary diagrams; inclusion of both elementary diagrams
  and triplet correlations.
}
\end{figure}

\begin{figure}[hbt]
{\includegraphics[width=0.4\textwidth,angle=-90]{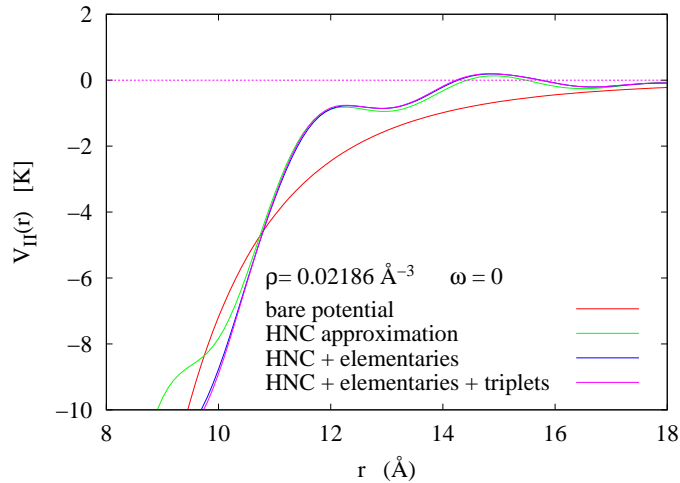}}
\caption[]{\label{fig:viif}
  As Fig.~\ref{fig:viiu}, but choosing an average frequency $\bar\omega=0$,
  Eq.~(\ref{eq:wind2}).
}
\end{figure}

The key result of our calculations is the effective interaction
(\ref{eq:Veff}). Figs. \ref{fig:viiu} and \ref{fig:viif} show the
HNC-EL results for the effective interaction, using the two versions
(\ref{eq:wind1}) and (\ref{eq:wind2}) of the induced interaction
defined above.  Hydrodynamic consistency was enforced as described in
section~\ref{ssec:HNCEL}. The figures also give information on the
importance of elementary diagram corrections and triplet
correlations. It is seen that the effect of triplet correlations is
almost invisible; elementary diagrams modify the effective interaction
only at short distances and can, therefore, have a visible influence
on the binding energy of the Mg dimer in its ground state.

The results of both versions of the induced interaction are
practically identical, the version using the zero-frequency response
function is only slightly more repulsive. We see in both cases two
slight local maxima at distances of about 12\,\AA\ and
14.4\,\AA. However these maxima are much too weak to permit a
metastable loosely bound state at large distances. We have
also calculated the bound state energies of the Mg dimer; there is
no evidence for an increased density of states in the energy
regime where the effective potential becomes oscillatory.

\subsection{PIMC simulation of Mg dimers}

We have also performed PIMC simulations for a Mg dimer surrounded by a
cluster of either 100 or 200 $^4$He atoms at a temperature of
$T=0.31$K, which is on the low end of the estimated temperature of
$^4$He droplets, $T=0.3 - 0.4$K, in typical experimental
conditions~\cite{toenniesAngChem04}.  Already such a small $^4$He
droplet might be able to form a ``protective'' $^4$He layer between
two Mg atoms, respectively, that was invoked in
Ref.~\onlinecite{przystawik} to explain why weakly bound Mg complexes
can be stable against collapse to the ground state.

Our search for a metastable Mg dimer state consists of three stages to
make the equilibration of the Mg degrees of freedom as gentle as
possible: ({\it i\/}) equilibrating $^4$He atoms with the two Mg atoms
held fixed at a distance of 10\AA\; ({\it ii\/}) releasing the Mg
atoms to allow them to equilibrate, but {\em without} the bare Mg-Mg
interaction -- hence only the effective phonon-mediated interaction
$V_{\rm ind}(r)$ is felt by the Mg.  Our HNC-EL calculation has shown
that $V_{\rm ind}(r)$ is oscillating with $r$ and thus has local
minima; ({\it iii\/}) switching on the bare Mg-Mg interaction.

\begin{figure}[hbt]
\includegraphics[width=0.49\textwidth]{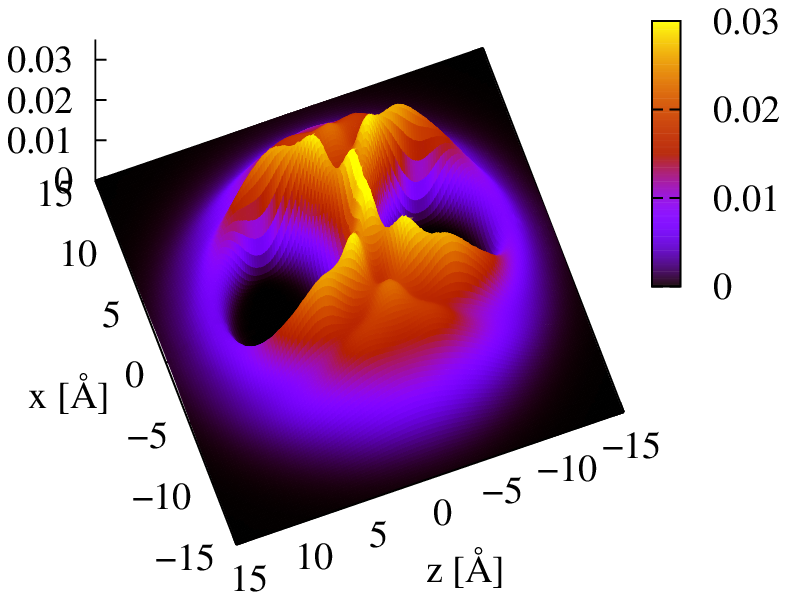}
\includegraphics[width=0.49\textwidth]{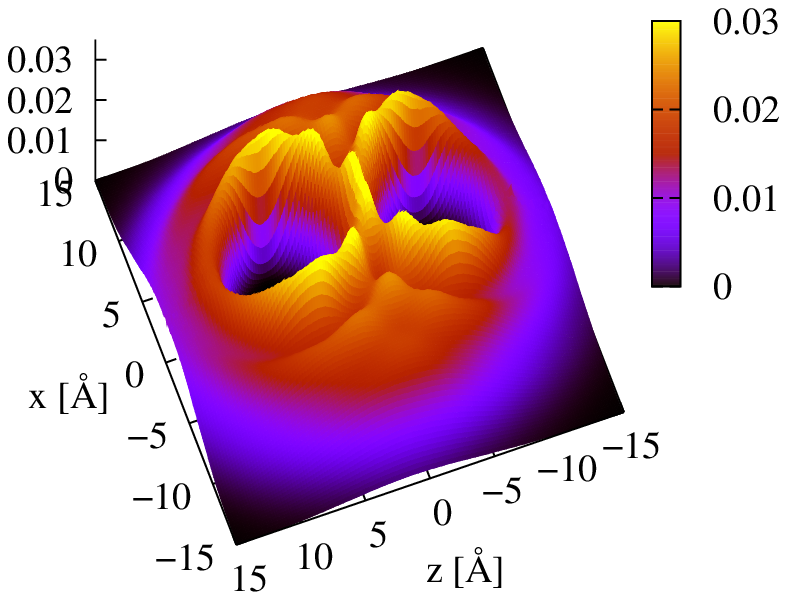}
\caption[]{\label{fig:3001}
  $^4$He density $\rho(\rvec)$ in the coordinate frame defined by
  the instantaneous Mg-Mg axis for a cluster of $N=100$ (left)
  and $N=200$ (right) $^4$He atoms.
}
\end{figure}

In stage ({\it ii\/}), the two Mg atoms indeed stay well separated
by approximately 10\,\AA .  Hence the local minima of just the phonon-mediated
effective interaction $V_{\rm ind}(r)$ indeed support bound states of
a well-separated Mg pair.  Note that this is a metastable state since the
ground state would correspond to the Mg atoms coalescing to the same
spot since there is no bare Mg-Mg interaction preventing this.
Fig.~\ref{fig:3001} shows the $^4$He density with respect to the Mg-Mg
axis for $N=100$ and $N=200$.

The $^4$He density is not zero between the two Mg atoms in the dimer;
these are separated by a liquid layer of helium.  Note that this result
also demonstrates that PIMC simulations can sample a metastable state,
where we take advantage that PIMC does not necessarily  ergodically
sample the full configuration space in the presence of a
high reaction barrier.  For example, we have observed such a
metastable state for electronically excited Rb on the surface of
helium by performing a ``vertical Monte Carlo'' transition from the
$\Sigma_{1/2}$ to the $\Pi_{1/2}$ and $\Pi_{3/2}$ states~\cite{leinoJCP08,leinoJCP11},
in agreement with experiment~\cite{auboeckPRL08}.

However, these simulations obtained by non-ergodic sampling do not yet
provide information on the height of barrier for the two Mg atoms to
coalesce to the same spot, {\em i.e.\/} whether the He layer between
two Mg atoms prevents the formation of Mg$_2$ in its ground state when
the bare Mg-Mg interaction is turned on.  To calculate the energy that
the Mg dimer has to overcome, we performed simulations as in stage
({\it i\/}), but fixing the Mg atoms at varying
distance $r$ and recorded the total
energy $E$ as function of $r$.  This corresponds to a Born-Oppenheimer
approximation for the He surrounding the Mg atoms, where the
Mg atoms are effectively assumed to have infinite mass.  In HNC-EL, the
proper bare mass of Mg is used.  A comparison between
$\tilde w_{\rm ind}(k)$, Eq. (\ref{eq:wind1}), in Fig.\ref{fig:viiu}
and $\tilde w_{\rm ind}'(k)$, Eq. (\ref{eq:wind2}), in Fig. \ref{fig:viif}
shows that, at least for the HNC-EL induced interaction, the
mass of Mg has only a small influence.

\begin{figure}[hbt]
\centerline{\includegraphics[width=0.7\textwidth]{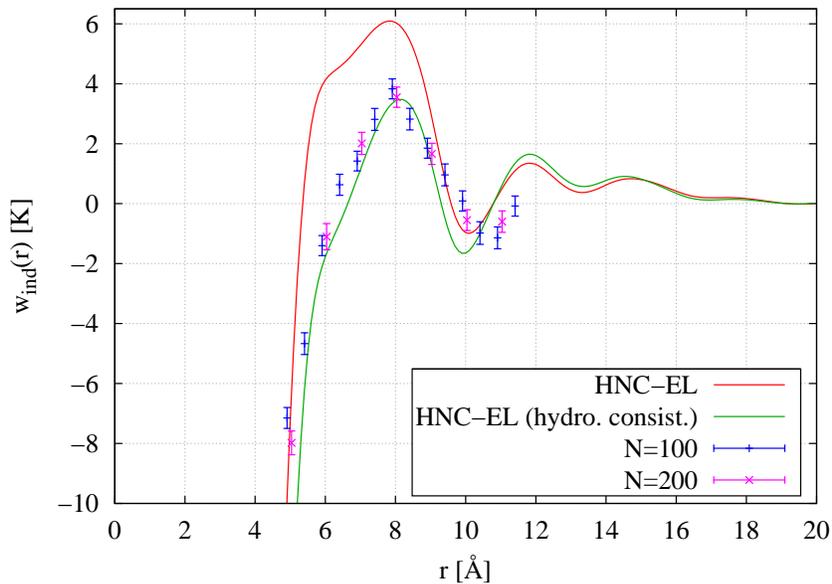}}
\caption[]{\label{fig:wind}
  Comparison of the {\em induced} Mg-Mg potential between HNC-EL for bulk $^4$He
  (red/green: without/with hydrodynamic consistency)
  and PIMC at $T=0.31$K, for clusters of $N=100$ (blue)
  and $N=200$ (pink) $^4$He atoms.
  The latter curve was obtained by subtracting X K and Y K from
  the total energy obtained by PIMC.
}
\end{figure}

The energy $E(r)$
is shown in Fig.~\ref{fig:wind}, together with the induced Mg-Mg
potential calculated by HNC-EL according to Eq.~(\ref{eq:wind1}).  The
agreement between the HNC-EL results and the PIMC calculations is
excellent if hydrodynamic consistency is ensured (see
section~\ref{ssec:HNCEL}, considering that very different approximations
are made for the HNC-EL calculations (variational wave function,
approximation for elementary diagrams) and PIMC calculations
(Born-Oppenheimer approximation), respectively.

\begin{figure}[hbt]
\centerline{\includegraphics[width=0.7\textwidth]{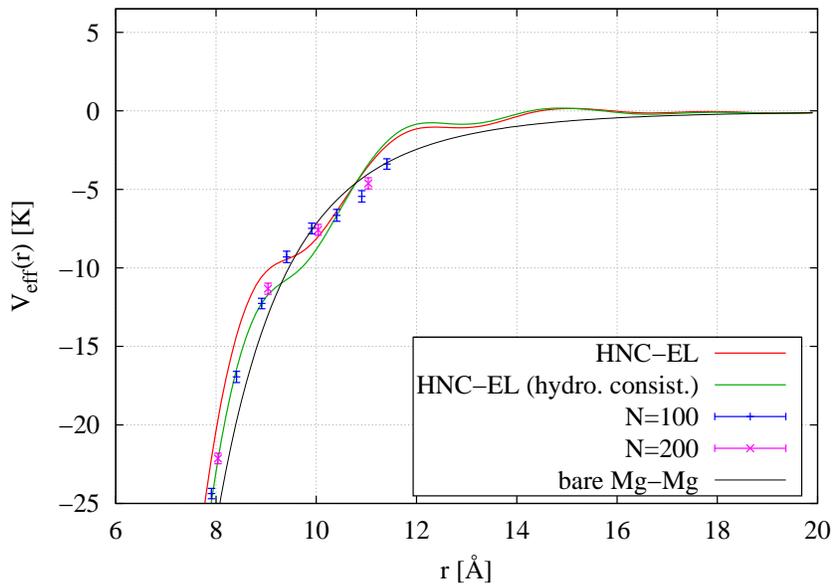}}
\caption[]{\label{fig:veff} Same as Fig.~\ref{fig:wind}, but for the
  effective Mg-Mg potential $V_{\rm eff}(r)$, i.e. adding the bare
  interaction according to Eq.~(\ref{eq:veff}).  }
\end{figure}

In Fig.~\ref{fig:veff}, we add the bare Mg-Mg interaction to $V_{\rm ind}(r)$
to obtain the full interaction $V_{\rm eff}(r)$ felt by a pair of Mg atoms.
The tenuous local minimum of the He-induced interaction clearly has a barrier
much too small to survive when the bare Mg-Mg interaction is added.

In stage ({\it iii\/}) in our PIMC investigation, we finally turn on
the bare Mg-Mg interaction.  From the above calculation of the $V_{\rm
  eff}(r)$ in Born-Oppenheimer approximation we expect that the Mg
pair will equilibrate from the weakly-bound metastable state observed
without the bare Mg-Mg interaction in stage ({\it ii\/}) to a strongly
bound state.  This is indeed what happens in every PIMC simulation,
trying 16 different initial configurations obtained in stage ({\it
  ii\/}).  We did not find evidence for a weakly bound dimer state,
when all interaction of the Hamiltonian are included in the
simulations.  Instead we always observe swift equilibration to the
dimer ground state.  While this PIMC result itself does not constitute
a proof that there is no weakly bound metastable dimer, the
Born-Oppenheimer potential energy together with the HNC-EL results
make a compelling case that such a state is highly unlikely.

We also investigated the (meta-)stability of Mg trimers.  Starting
with three Mg atoms situated at large distance from each other in a
triangle, PIMC simulations without the bare Mg-Mg interactions
equilibrated towards metastable states similar to the corresponding
simulations of Mg dimers, {\em i. e.\/} each Mg solvated by a
protective layer of He instead of the three Mg collapsing to the same
spot.  However, as for Mg dimers, turning on the bare Mg-Mg
interactions quickly lead to Mg$_3$ equilibrating towards the trimer
ground state.

\section{Conclusion}

The main goal of our work was to investigate the effective potential
barrier with two alternative and manifestly microscopic methods
theoretical methods.  We have used the variational HNC-EL method,
based on Jastrow-Feenberg wave functions which include correlations
between He atoms and between He and the molecule, which are absent in
DFT.  Due to the increased complexity due to correlations we have
restricted ourselves to Mg atoms in bulk $^4$He.  Additionally, we
used PIMC to simulate Mg atoms in bulk $^4$He and in small droplets of
$^4$He.  We found very good agreement of the solvation structure
around an Mg atoms, described by the static structure function.  We
also confirmed that Mg atoms reside inside He sufficiently large
droplets, but only barely -- slightly elevating the temperature leads
to significant population of surface states.

 We have also derived approximate Mg-Mg-He triplet correlations based
 on the HNC-EL results for a single impurity, and from that the
 induced Mg-Mg interaction, {\em i.e.\/} the effective interaction
 mediated by excitation of the surrounding He medium. Since PIMC
 simulations of two widely separated Mg atoms in bulk He would be
 computationally very demanding, we opted for two Mg atoms in small
 $^4$He clusters, of $N=100$ and 200 atoms.  Furthermore, simulations
 where the two Mg atoms can move freely among the He atoms very
 quickly find the stable ground state of a closely bound Mg dimer,
 despite starting the simulations with a large initial separation
 between the Mg atoms.  Therefore, we had to use a Born-Oppenheimer
 approximation, where the heavier Mg atoms are held fixed at distance
 $R$ while the He atoms are allowed to move in the simulation.
 Repeating the simulation for many different $R$ maps out the
 effective interaction between the Mg atoms.  Since Mg atoms are only
 six times heavier than He atoms, we should expect an effective
 interaction biased by the Born-Oppenheimer approximation; we note
 that the DFT estimates of Ref. \onlinecite{hernandoPRB08b} also used
 this Born-Oppenheimer approximation.

The most severe potential inaccuracy of the Born-Oppenheimer
approximation is, however not because of the mass of the Mg atoms as
discussed above, but due to the dynamic coupling of the zero-point
motion of the dimer to the surrounding helium through hydrodynamic
backflow. In fact, one would expect that this backflow effect is
different dependent on whether the dimer is in an excited state or in
the ground state. A quantitative assessment of the effect is very
difficult and the theoretical tools are presently not
available. We have therefore estimated the effective mass using the
methods of Ref.~\cite{SKJLTP} and found a value of $m^*/m \approx
2$. This mass enhancement is still far from what would be needed to
generate a sufficiently long-lived metastable state in the relative
potential minimum around 10\,\AA.

We found that neither the HNC-EL results not the PIMC results for the
induced Mg-Mg interaction supports a local minimum in the total
effective interaction consisting of bare and induced interaction.  The
results for the two methods agree very well, although the agreement
may well accidentally, considering the different assumptions and
approximations made in the PIMC simulations (Born-Oppenheimer
approximation, He droplets up to $N=200$ as opposed to bulk He).
Thus, using two completely different methods, we find no evidence for
a metastable Mg ``bubble foam'' in helium.  We did not consider the
effects of an angular momentum barrier, since its hard to argue how a
spinning Mg dimer would not decay on a time scale of $\mu$s, and how
to construct an angular momentum barrier for ``bubble foam'' made of
many Mg atoms.  There is the possibility that metastable Mg clusters
beyond dimers are stable, but at least for Mg trimers our PIMC
simulations seem to rule that out.

Since spectroscopic experiments of Mg clusters in $^4$He droplets were
interpreted as evidence of a meta-stable ``bubble foam'' of Mg atoms
in the helium matrix\cite{mgfoam_expt}, a conclusive theoretical
understanding of these spectra is still missing.

\begin{acknowledgments}

  This work was supported, in part, by the College of Arts and
  Sciences, University at Buffalo SUNY, and the Austrian Science Fund
  Projects P21264 and I602 (to EK) and P23535 (to REZ).  Discussions
  with Yaroslav Lutsyshyn are also acknowledged.

\end{acknowledgments}

\bibliography{papers,my,bec,ocshehy}
\bibliographystyle{apsrev4-1}

\end{document}